\documentclass{webofc}
\usepackage[varg]{txfonts}   
\usepackage{xcolor}

\newcommand{\GE}{\mbox{${G_{_{\!E}}}$}}
\newcommand{\GEsqr}{\mbox{${G_{_{\!E}}^2}$}}
\newcommand{\GEorM}{\mbox{${G_{_{\!E/M}}}$}}
\newcommand{\GM}{\mbox{${G_{_{\!M}}}$}}
\newcommand{\GMsqr}{\mbox{${G_{_{\!M}}^2}$}}
\newcommand{\GEp}{\mbox{${G_{_{\!E}}^{p}}$}}
\newcommand{\GMp}{\mbox{${G_{_{\!M}}^{p}}$}}
\newcommand{\dd}[2]{\frac{\mathrm{d}#1}{\mathrm{d}#2}}

\begin{document}
\title{The MESA physics program}
\author{\firstname{Sören}       \lastname{Schlimme}     \inst{1}    \fnsep\thanks{\email{schlimme@uni-mainz.de}} \and
        \firstname{Kurt}        \lastname{Aulenbacher}  \inst{1,2,3}\and
        \firstname{Sebastian}   \lastname{Baunack}      \inst{1}    \and
        \firstname{Niklaus}     \lastname{Berger}       \inst{1,2}  \and
        \firstname{Achim}       \lastname{Denig}        \inst{1,2,3}\and
        \firstname{Luca}        \lastname{Doria}        \inst{1}    \and
        \firstname{Alfons}      \lastname{Khoukaz}      \inst{4}    \and
        \firstname{Frank}       \lastname{Maas}         \inst{1,2,3}\and        
        \firstname{Harald}      \lastname{Merkel}       \inst{1,2}  \and
        \firstname{Concettina}  \lastname{Sfienti}      \inst{1,2,3}\and
        \firstname{Michaela}    \lastname{Thiel}        \inst{1}
}
\institute{
Institut für Kernphysik, Johannes Gutenberg-Universität, D-55099 Mainz, Germany \and 
PRISMA$^+$ Cluster of Excellence, Johannes Gutenberg-Universität, D-55099 Mainz, Germany \and 
Helmholtz Institute Mainz, GSI Helmholtzzentrum für Schwerionenforschung, Darmstadt, Johannes Gutenberg-Universität, D-55099 Mainz, Germany \and
Institut für Kernphysik, Westfälische Wilhelms-Universität, D-48149 Münster, Germany
}
\abstract{%
In the recent past, a comprehensive experimental program has been worked out at the Mainz Energy-Recovery Superconducting Accelerator, MESA, at the Institute of Nuclear Physics in Mainz.
MESA is a high-intensity, low-energy electron accelerator presently under construction and will thereby provide great opportunities to perform a new generation of high-precision scattering experiments. 
The versatile MAGIX experiment will use MESA's innovative energy recovery technique, with a science focus on the study of hadron structure and few-body systems, dark sector searches, and investigations into reactions pertinent to nuclear astrophysics.
An external beam line will supply spin-polarized electrons to the P2 experiment, enabling the performance of sensitive tests of the Standard Model through parity-violating electron scattering. 
The DarkMESA beam dump experiment, situated behind P2, is dedicated to the search for light dark matter particles. 
}
\maketitle
%
\section{Introduction}\label{sec:Intro}
%
In an effort to address fundamental questions in nuclear and particle physics, a new accelerator called MESA has been developed \cite{MESA_portrait,Hug_MESA2019}. This state-of-the-art facility has been carefully designed to enable a new era of precision scattering experiments at low energy and promises to further deepen our understanding of the subatomic world.

MESA is currently being installed underground at the Institute of Nuclear Physics in Mainz, situated adjacent to the long-standing and successful MAMI accelerator, which has been in operation for several decades \cite{Dehn_MAMI25}.
Electrons are provided by a photo-gun and pre-accelerated to $5\,\mathrm{MeV}$ with a beam time structure of 1.3 GHz, resulting in a quasi continuous wave beam. The core of the accelerator comprises two cryomodules, each housing two superconducting cavities. With each pass through these cryomodules, electrons can acquire an additional $25\,\mathrm{MeV}$ of energy. MESA has the capability to operate in two distinct modes.

In Extracted-Beam (EB) mode, after three circulations, spin-polarized electrons are supplied with a beam current of up to $150\,\mu\mathrm{A}$ and a beam energy of $155\,\mathrm{MeV}$. Subsequently, these electrons are directed towards the P2 experiment, tailored for parity violation experiments, before being safely stopped in a high-power beam dump.
In line with the beam dump but outside the accelerator hall is the beam dump experiment DarkMESA located. Dedicated to the search for light dark matter particles, which could potentially be produced inside the dump, this experiment benefits from the extensive radiation shielding provided by the dump. 

In Energy-Recovery Linac (ERL) mode, the beam undergoes initial acceleration up to $105\,\mathrm{MeV}$ before being redirected back to the cavities with a 180-degree phase shift following the MAGIX experiment location. Consequently, the electrons experience deceleration, resulting in the restoration of energy to the accelerating structures -- again up to $25\,\mathrm{MeV}$ for each pass. At $5\,\mathrm{MeV}$, the beam is directed into a compact beam dump located a considerable distance from the MAGIX site, ensuring minimal background interference during experimentation.
The ERL mode offers highly energy-efficient acceleration, enabling the attainment of substantial beam currents of $1000\,\mu\mathrm A$ or more.
To achieve this, the disturbance of the beam when crossing the target must be kept to a minimum. Consequently, the primary target of MAGIX will be a thin, windowless gas jet target \cite{Schlimme_JetAtA1}. Despite the overall competitiveness of luminosity resulting from the large beam current (up to $10^{35}\,\mathrm{cm^{-2}s^{-1}}$), the impact of undesired effects induced by the target -- such as energy loss spreading of particles, multiple scattering, or background from a target window -- becomes negligible. 

Set to begin in mid-2025, the experiments cover a wide range of research projects, as detailed in the following sections.
\begin{figure}[tb]
	\begin{center}
		\includegraphics[trim=0 300 0 200, width=\textwidth]{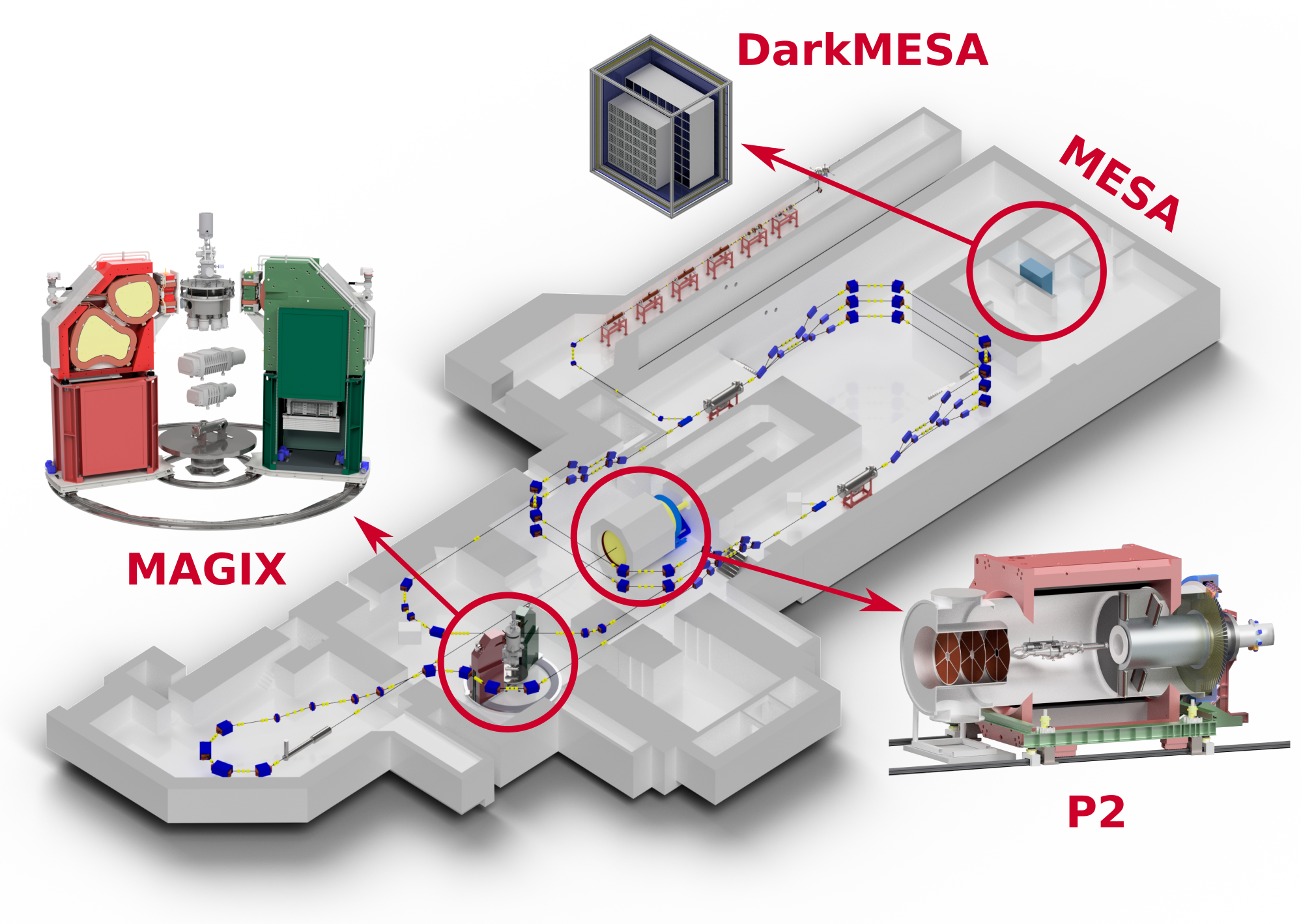}
	\end{center}
	\caption{Schematic overview of MESA, highlighting its three experiments: MAGIX, P2, and DarkMESA.}
	\label{fig:MESA_floor}
\end{figure}

\section{High-precision electron scattering experiments at MAGIX}\label{sec:MAGIX}
Central components of the multi-purpose MAinz Gas Injection Target EXperiment MAGIX are the gas jet target, two high-resolution magnetic spectrometers as well as supplementary recoil detectors. When deemed advantageous, solid state targets can be employed in EB mode of MESA as well.
The electron beam interacts with the target in the scattering chamber, which is directly connected to the beam pipe without any windows. 
Particles resulting from interactions at the target can enter the acceptance of either of the two identical magnetic spectrometers, where they are focused to focal plane detectors, which consist of a time projection chamber and a trigger veto system for each spectrometer. To detect recoil particles of the gas jet in addition to the scattered electrons, a silicon strip detector based recoil detector system can be placed in the scattering chamber. 
Given the relatively low beam energy, the energy of the scattered particles is limited; thus, the entire system has been optimized to minimize particle interactions prior to detection.

This setup is exceptionally well-suited for high-precision examinations of nuclear structure and dynamics at low energies, as well as for particle searches using the missing mass technique.

\subsection*{Structure of Nucleons and Nuclei}\label{sec:MAGIX:Structure}
Electron scattering experiments have been crucial in advancing our understanding of the internal structure of hadrons and probing the distribution of their charge and magnetization. One noteworthy example is the examination of the $Q^2$ dependence of the electric ($\GE$) and magnetic ($\GM$) Sachs form factors (FFs) of both protons and neutrons, which can be obtained from unpolarized cross-section measurements of the elastic scattering process,
\begin{equation*}
    \left(\dd{\sigma}{\Omega}\right) = \left(\dd{\sigma}{\Omega}\right)_{\text{Mott}} \cdot \frac 1{\varepsilon\,(1+\tau)}\left(\varepsilon \,\GEsqr(Q^2)+\tau \,\GMsqr(Q^2)\right), \qquad \tau=\frac{Q^2}{4M^2},
\end{equation*}
by performing a Rosenbluth separation \cite{JCB_pFF2014}. Here, $\varepsilon$ is the polarization of the exchanged virtual photon, $M$ the nucleon mass.
$\GE$ and $\GM$ are directly related to the distributions of charge and magnetization inside the nucleons, respectively. 
Data at small momentum transfers are of particular interest, as the root-mean-square charge and magnetic radii can be extracted from the slope of the FFs at zero momentum transfer,
\begin{equation*}
    \langle r_{\mathrm E/M}^2\rangle = - 6 \hbar^2\frac{1}{\GEorM(0)}\left.\dd{\GEorM}{Q^2}\right|_{Q^2=0}
\end{equation*}
A significant discrepancy in the measurement of the proton radius, a synonym for its charge radius $\sqrt{r_{\mathrm E}^2}$, using the two distinct methods of electron scattering and muonic hydrogen spectroscopy, has given rise to the famous proton radius puzzle. This discrepancy remains a subject of active debate and investigation to this day \cite{Walcher:2023qyk}. Experiments like MUSE, AMBER, PRAD-II, and ULQ$^2$ are set to provide additional scattering data, employing different approaches to clarify this matter.

At MAGIX, a dedicated dataset will be collected at very low momentum transfers, down to $Q^2\sim 10^{-5}\,\mathrm{GeV^2}/c^2$. 
The experiment is expected to benefit greatly from the use of the gas jet target. Unlike the challenge of background subtraction encountered in the analysis of previous MAMI-A1 data \cite{JCB_pFF2014}, the upcoming measurement is expected to be virtually free of background. Furthermore, the point-like intersection of the electron and hydrogen beams simplifies track reconstruction, and the use of the relatively thin target will significantly reduce effects from external radiation \cite{Schlimme_JetAtA1,Yimin}.

\begin{figure}[htbp]
	\begin{center}
		\includegraphics[width=0.7\textwidth]{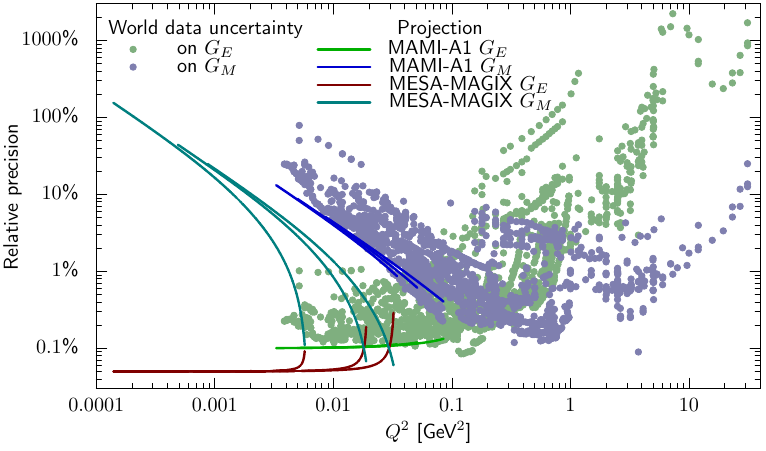}
	\end{center}
	\caption{
        Illustration of the sensitivity of world cross-section data to $\GEp$ and $\GMp$. This sensitivity is calculated by translating individual cross-section measurement uncertainties to the uncertainty of the respective FF, assuming that the other FF is known perfectly. At large $Q^2$, the $\tau$-factor amplifies the contribution of $\GMp$ to the cross-section, making the data primarily sensitive to $\GMp$. Conversely, at low $Q^2$, $\GEp$ dominates the cross-section except for extreme backward scattering. Figure from \cite{JCB_rp}. PRad data is only sensitive to $\GEp$ and therefore omitted.
        }
	\label{fig:Gp_Projected}
\end{figure}

Moreover, with the small beam energies available at MESA, the experiment aims to achieve an order of magnitude improvement in the precision of the magnetic form factor in the region crucial for determining the magnetic radius and the Zemach radius, establishing another link to atomic physics, see Fig.~\ref{fig:Gp_Projected}.
The versatility of the jet target, compatible with most gases \cite{Grieser}, provides the opportunity to extend similar structure studies to other nuclei.

\subsection*{Few-Body Systems}\label{sec:MAGIX:FewBody}

{\it Inelastic} electron scattering can provide valuable insights into nuclear dynamics, playing a crucial role in testing theoretical calculations derived from any Effective Field Theory (EFT) against high-precision electron scattering data. This process is essential for refining our understanding and modeling of strongly interacting systems. The perturbative nature of the electromagnetic probe enables a clean comparison between theory and experiment~\cite{Bacca14}, fostering a robust interplay between the two. Moreover, the presence of external fields requires the inclusion of electroweak current operators in electron scattering observables, a requirement that should be consistently addressed within an EFT framework.

A recent example of this interplay is the investigation of the monopole transition form factor of $^{4}$He from the $0^+$ ground state to the resonant $0^+$ state. A measurement at MAMI-A1 with a precision surpassing that of older data sets strongly contradicted predictions based on state-of-the-art calculations derived within chiral effective field theory and well tested on a variety of other observables \cite{Kegel23}. 
The MAMI data have stimulated extensive theoretical discussion, and a detailed evaluation of the impact is still in progress.

In accordance with this approach, the capabilities of MAGIX will be harnessed to conduct dedicated measurements in the few-body sector.
The objective is to test or calibrate low-energy nuclear theories within the ideally matching low-energy regime of MESA, where they are supposed to work best. A measurement of the aforementioned monopole transition form factor at very low momentum transfers is planned.
Additionally, the investigation will encompass the inclusive $^4$He(e,e')X reaction, with a particular focus on the longitudinal response function $R_L$, where significant three-nucleon-force effects are predicted but remain untested. The consistent windowless design of the thin MAGIX gas-jet target even allows for the detection of low-energy recoil particles, such as protons or deuterons. This is particularly relevant as it facilitates measurements where a knocked-out fragment of a nucleus is detected in coincidence with the scattered electron. Such exclusive reactions can provide more informative data compared to inclusive ones. 

In addition, a series of measurements on deuteron electrodisintegration will be conducted at low energy and forward angles. The aim is to enhance the precision of nuclear structure corrections in muonic deuterium resulting from the two-photon exchange process. 
At a higher mass scale, there are plans to conduct a precise measurement of the monopole transition form factor to the Hoyle state -- the $0^+$ excited state of the $^{12}$C nucleus. This measurement will be compared to calculations treating the Hoyle state as a system of three $\alpha$-particles interacting with forces derived within a cluster effective field theory. Notably, this excited state plays an extraordinary role in the astrophysical triple-$\alpha$ carbon fusion process.

\subsection*{Study of reaction cross-sections of astrophysical interest}\label{sec:MAGIX:Astro}
%
Stellar fusion processes are essential for building models that explain the abundance of elements in the universe. 
Precise cross-section estimates, particularly for nucleon and light nuclei capture, are crucial for this modeling. Directly measuring these cross-sections at the low energies relevant to stellar burning is extremely challenging since the cross-sections at these energies are extremely small. Consequently, measurements are dominated by background and may be inaccessible in the laboratory, especially when dealing with an unstable nucleus.

\begin{figure}[b]
    \begin{center}
    \begin{minipage}{.2\textwidth}
    \includegraphics[width=\textwidth]%
    {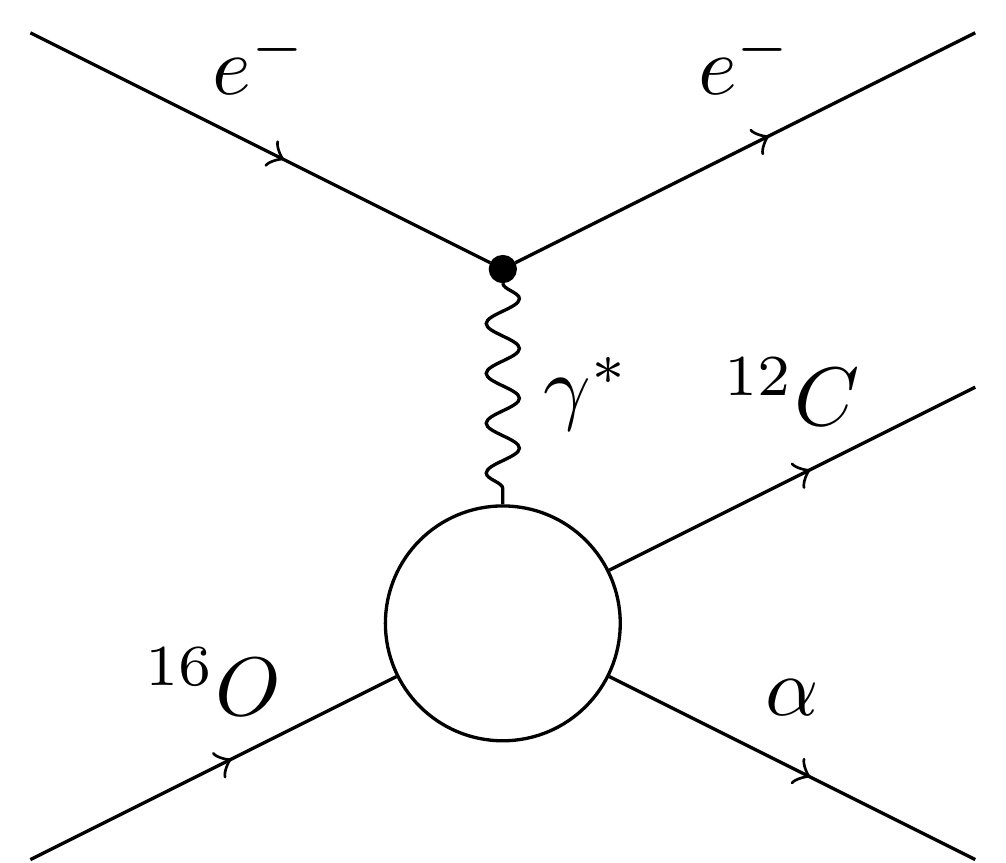} 
    \end{minipage}\qquad
    \begin{minipage}{.4\textwidth}
    \includegraphics[width=\textwidth]%
    {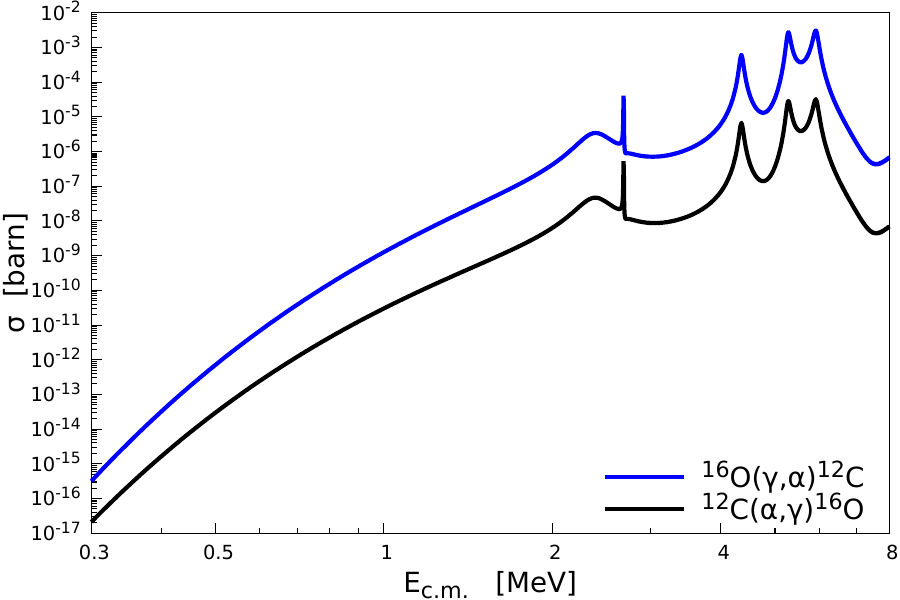} 
    \end{minipage}
    \end{center}
    \vspace{-0.5cm}
    \caption{Illustration of the photo-dissociation technique to infer radiative-capture cross-sections in the case of $\rm ^{12}C(\alpha,\gamma)^{16}O$.
    Left: a high-intensity electron beam induces the dissociation of $\rm ^{16}O$ into $\rm ^{12}C$ and an $\alpha$ particle by the exchange of a photon.
    Right: the cross-section for the radiative capture $\rm ^{12}C(\alpha,\gamma)^{16}O$ (black line) drops dramatically at very low energies. Between the lowest until now reached center-of-mass energy, $1\,\mathrm{MeV}$, and the point of helium burning at $0.3\,\mathrm{MeV}$, the cross-section drops by nearly six orders of magnitude, and an extrapolation of existing data to the relevant energy range introduces significant theory dependence. The cross-section for the photo-dissociation process (blue line) is significantly larger \cite{Lunkenheimer_Astro16O}.
    }
    \label{fig:dissociation}
\end{figure}

At MAGIX, the objective is to deduce radiative-capture cross-sections for several key reactions at low energy using an indirect technique. Photo-dissociation reactions, such as $\rm (\gamma,\alpha)$, 
will be induced by MESA's high-intensity electron beam. This is achieved using quasi-real photons, i.e., photons with a very small four-momentum transfer compared to the scale set by the nuclear form factor \cite{Friscic}, see Fig.~\ref{fig:dissociation} (left panel). By measuring electrons with the high-resolution MAGIX spectrometers in coincidence with the recoil particles, a significant reduction in background contributions can be achieved. Since photo-dissociation and radiative capture represent time-reversed processes, their cross-sections are directly correlated. The radiative-capture cross-section of interest can then be inferred from the measured one through a detailed balance. 

\begin{figure}[t]
    \begin{center}
    \includegraphics[width=\textwidth]{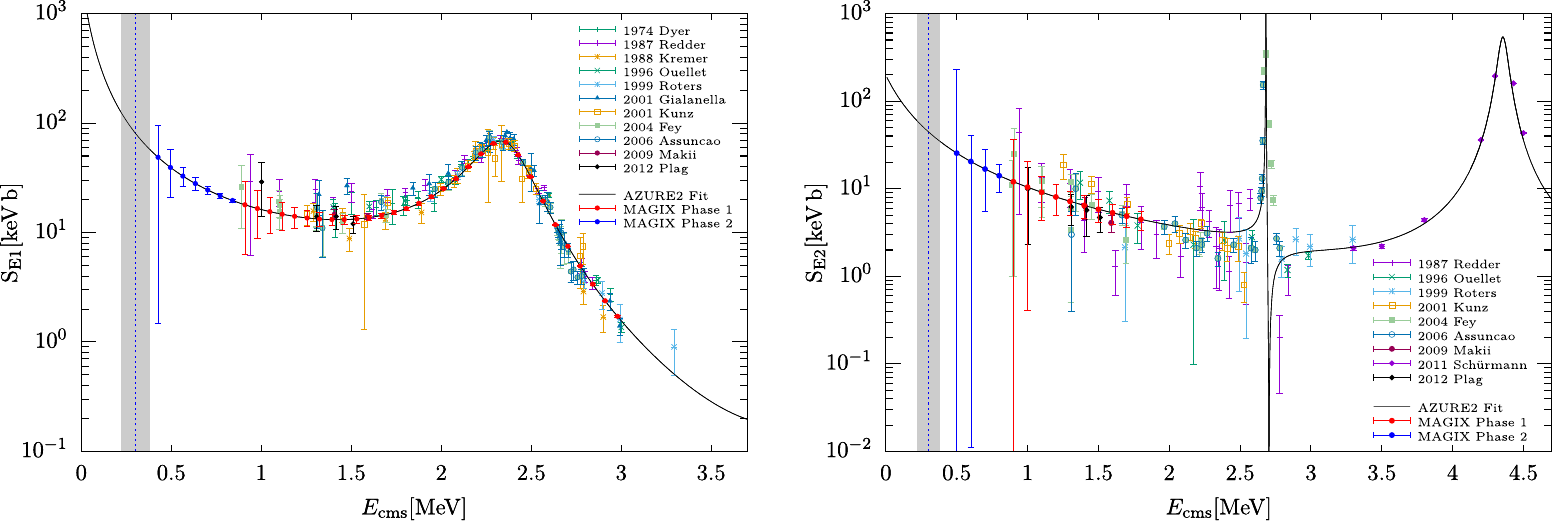} 
    \end{center}
    \caption{$\rm E1$ (left) and $\rm E2$ (right) contributions to the astrophysical $S$ factor (non-trivial part of the cross-section characterizing the unknown nuclear physics) of the $\rm ^{12}C(\alpha,\gamma)^{16}O$ radiative capture \cite{Lunkenheimer_Astro16O}. In addition to existing data (various points) an R-matrix estimate is also shown (AZURE2 Fit; black solid line).
    The red and blue points show simulations of the MAGIX Phases 1 and 2, respectively. The Gamow peak (representing the energy where reactions are most probable) is indicated by the dashed blue line and its width by the gray band.}
    \label{fig:AstroS}
\end{figure}

In Fig.~\ref{fig:dissociation} (right panel), this connection is illustrated for the famous radiative alpha particle capture by a carbon nucleus. This reaction holds significant importance in describing nucleosynthesis during stellar burning, as it is essential for determining the $\rm ^{12}C / ^{16}O$ ratio. This ratio, in turn, plays a dominant role in subsequent processes and thereby influences nuclear abundances significantly.
According to simulations, the MAGIX setup is anticipated to enable the measurement of astrophysical cross-sections for center-of-mass energies above $1\,\mathrm{MeV}$ with a precision competitive to existing data ({\it MAGIX Phase 1} in Fig.~\ref{fig:AstroS}). 
The inclusion of a zero-degree tagger to detect electrons scattered at very forward angles extends the capability to even lower energies, bringing the data
very close to the Gamow peak of this reaction ({\it MAGIX Phase 2}).

%
Other opportunities to use this technique at MAGIX include reactions like $\rm ^{24}Mg(\alpha,\gamma) ^{28}Si$ and $\rm ^{24}Mg(\gamma,\alpha)^{20}Ne$ (relevant during the silicon-burning phase in the late evolution of massive stars prior to core collapse), radiative proton capture reactions like $\rm ^{15}N(p,\gamma)^{16}O$ (key process in CNO cycle) and $\rm d(p,\gamma)^3He$ (rate of deuterium burning, primordial deuterium abundance), 
or radiative neutron capture reactions like $\rm ^{86}Rb(n,\gamma)^{87}Rb$ ($\rm ^{86}Rb$ being a branching point in the s-process within massive AGB stars)
and $\rm ^{204}Tl(n,\gamma)^{205}Tl$ ($\rm ^{204}Tl$: branching point in the s-process). For the reactions including a neutron, a dedicated neutron detector is required in addition to the basic MAGIX setup.

\subsection*{Dark Sector Searches}\label{sec:MAGIX:DSS}
The compelling astrophysical and cosmological evidence for the existence of Dark Matter (DM) has led to numerous experimental searches. However, an unambiguous signal for a DM candidate has not been found so far. Initially, the experiments have been targeting weakly interacting massive particles (WIMPs) and have been carried out either at high-energy colliders, or by searching for the recoil of the slowly moving WIMP particles in underground laboratories. With the lack of discovery of massive particles beyond the Standard Model (SM), the research focus has shifted in recent years towards searching for Light Dark Matter (LDM) candidates (below the GeV scale) and messenger particles of a hypothetical dark sector.

Assuming that these messenger particles are of vector nature (dark photons), they can undergo mixing with ordinary photons. It is believed that they couple to SM matter very weakly, characterized by a mixing parameter $\varepsilon=\sqrt{\alpha'/\alpha}$, with the dark and SM electromagnetic couplings $\alpha'$ and $\alpha$, respectively. The possibility of a dark photon to decay to DM depends on the masses of both the DM candidate and the dark photon: If the direct decay is kinematically feasible, the dark photon will decay {\it invisibly}. 
Otherwise, a massive dark photon will couple to SM particles in exactly the same way as an ordinary photon ({\it visible decay}).

\begin{figure}[b]
    \centering
    \includegraphics[width=\textwidth]{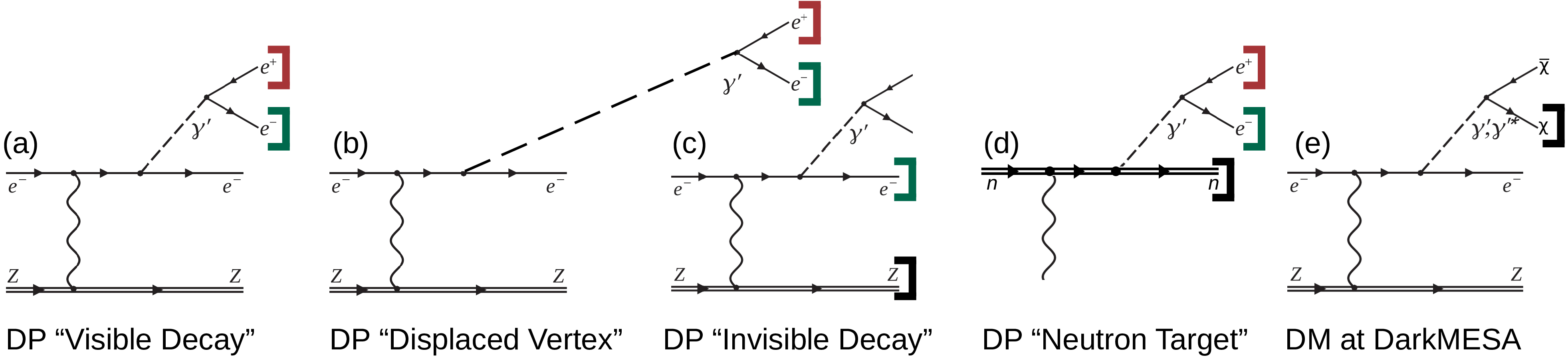}
    \caption{Dark sector searches will be performed at MAGIX using a high-intensity electron beam with a focus on dark photons (a)-(d). The sensitivity to various models will be exploited by using different reaction modes, see the text for details. 
    In addition, DarkMESA will be employed to explore the existence of light dark matter particles (e).
    }\label{fig:VariousDPMeasurements}
\end{figure}
\begin{figure}[b]
    \centering
    \includegraphics[width=0.42\textwidth]{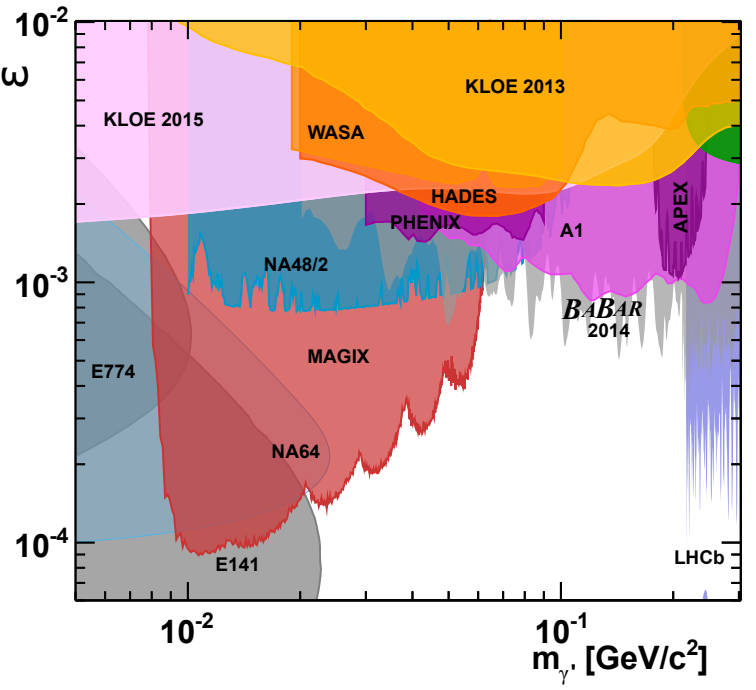}
    \includegraphics[width=0.57\textwidth]{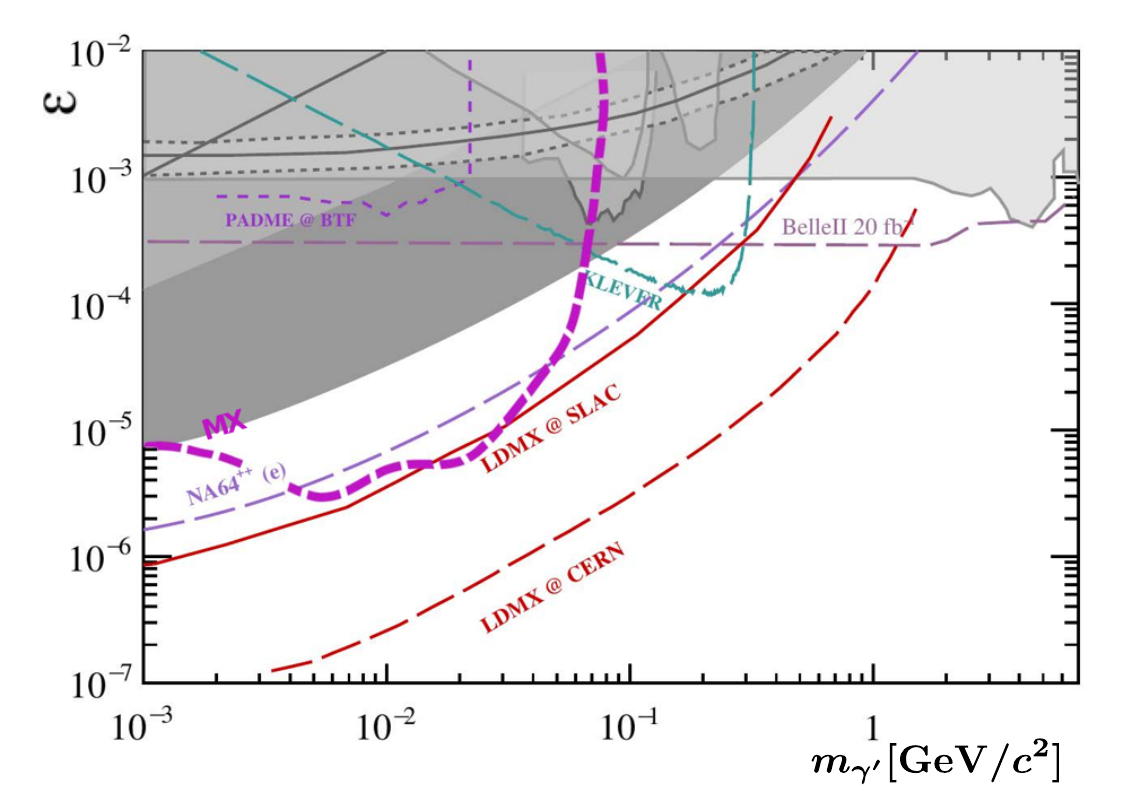}
    \caption{Left: The MAGIX sensitivity range for the visibly decaying dark photon using MESA beam energies up to $105\,\mathrm{MeV}$. Right: The range and sensitivity of the invisible decay experiment under consideration (labeled as MX) exhibit similarities to those of the visible decay channel at MAGIX. However, it is sensitive to different models.    }\label{fig:DPExclusionPlotsMX}
\end{figure}

World-class searches for dark photons are aimed at MAGIX, with investigations into both visible and invisible decay modes.
Fig.~\ref{fig:VariousDPMeasurements} provides a concise overview of the different experimental methods employed in this project.
The search for the visible decay of a dark photon will be conducted using MAGIX's two high-resolution magnetic spectrometers (a).
By positioning the target outside the acceptance range of the spectrometers' target length, it enables the exploration of low-$\varepsilon$ regions ("long-lived" dark photons) due to the significantly reduced QED background (b). 
Through the detection of the scattered electron and the recoiling nucleus, the presence of a dark photon can be revealed in the reconstructed missing invariant mass, irrespective of the dark photon's favored decay channels (c). 
In addition, there are plans to conduct a dedicated study of the dilepton production process on an effective neutron target, which has shown high sensitivity in previous studies \cite{Mommers:2024eeq} (d).
Experimental investigations focused on the search for dark matter and other exotic particles will be performed using the dedicated beam dump experiment DarkMESA (e).

Estimates for the sensitivity for the search for the visible decay of dark photons as well as in the invisible decay mode are shown in Fig.~\ref{fig:DPExclusionPlotsMX}.
%
\section{Parity violation experiments at P2}\label{sec:P2}
The primary objective of the P2 experiment \cite{Becker_P2} is to attain a highly precise measurement of the weak mixing angle at low energy as a sensitive test of the Standard Model. This will be achieved through the examination of the parity-violating asymmetry in elastic electron-proton scattering. 
Nevertheless, depending on the chosen target and kinematics, the experiment opens avenues to explore various other compelling physical quantities.

P2 makes full use of the high beam polarization of $85\,\%$ that will be available at MESA.
The experimental setup includes a superconducting solenoid, a ring of integrating Cherenkov detectors, supplementary tracking detectors, an additional backward angle detector, and it offers flexibility in selecting either a liquid hydrogen or solid-state target.
Longitudinally polarized beam electrons scatter on the unpolarized target. 
The solenoid's magnetic field guides elastically scattered electrons towards the detectors. The key measurement is the parity-violating cross-section asymmetry, determined by regularly altering the beam helicity. The tracking detector is used to measure the mean four-momentum transfer $\langle Q^2\rangle$. 
\subsection*{Precision measurement of the weak mixing angle}\label{sec:P2:sin2}
\begin{figure}[h]
    \centering
    \sidecaption
    \includegraphics[width=0.5\columnwidth]{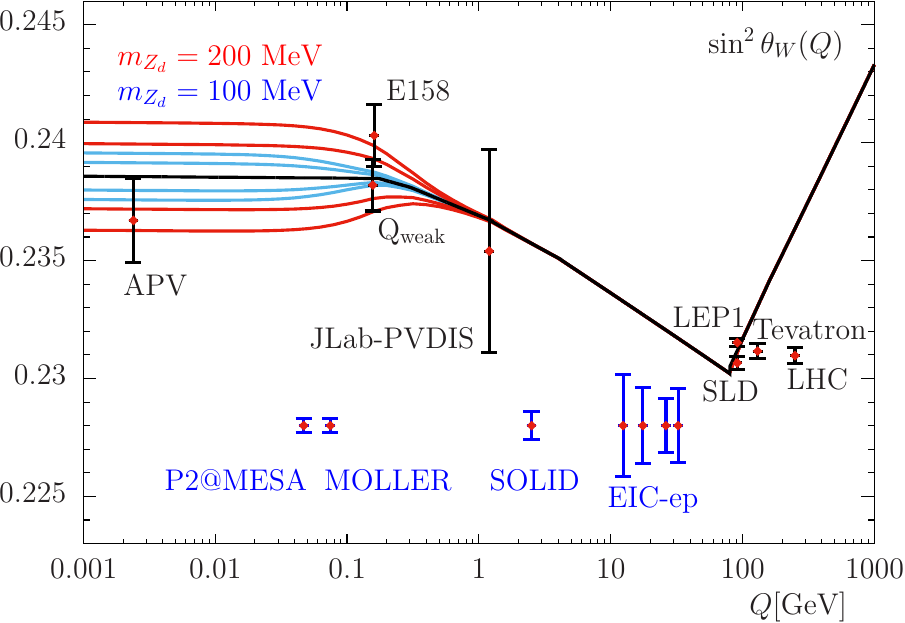} 
    \caption{
    Scale dependence of $\sin^2 \theta_W$ together with completed (black error bars) and planned (blue error bars) experimental measurements. The running of the effective weak mixing angle is modified if a dark $Z$-boson ($Z_d$) exists. Two scenarios from \cite{Davoudiasl:2014kua} are shown for $Z_d$ masses of $m_{Z_d} = 100\,\mathrm{MeV}$ and $200\,\mathrm{MeV}$.
    }
    \label{fig:P2_MixingAngle}
\end{figure}

The weak mixing angle $\theta_W$ is one of the fundamental parameters of the SM characterizing the mixing between the neutral weak force and the electromagnetic force in the electroweak theory of particle physics. It has been measured with great precision at the $Z$-pole. Quantum corrections make the weak mixing angle depend on the energy scale. 
A precise determination at low scales is therefore sensitive to contributions of new physics beyond the SM which can change the running of $\sin^2 \theta_W$.  As an example, a class of models extending the SM by postulating the existence of a light dark $Z$-boson with parity-violating couplings is specific to low energies and complementary to LHC (large hadron collider) searches for new physics. Deviations from the SM might only be noticeable at lower energies, as depicted in Fig.~\ref{fig:P2_MixingAngle}. The figure provides an overview of current and planned experiments, along with theoretical predictions illustrating the running of the weak mixing angle.

To determine $\sin^2 \theta_W$, the parity-violating cross-section asymmetry $A_{\mathrm{PV}}$ for left-handed and right-handed electrons will be measured in electron-proton scattering. It is connected to the weak charge of the proton $Q_{\mathrm W}$ through
$A_{\mathrm{PV}}=\frac{\sigma_{\mathrm{R}}-\sigma_{\mathrm{L}}}{\sigma_{\mathrm{R}}+\sigma_{\mathrm{L}}} 
  =\frac{G_{\mathrm F} Q^2}{4\sqrt 2 \pi \alpha_{\mathrm{em}}}\left(Q_{\mathrm W}-F(Q^2)\right)$,
where $G_{\mathrm F}$ is the Fermi constant, $\alpha_{\mathrm{em}}$ the fine structure constant, and $Q^2$ the squared four-momentum transfer, setting
the scale. The nucleon structure term $F(Q^2)$ includes contributions from the electromagnetic nucleon form factors, the axial form factor of the proton, and the strangeness contribution to the proton form factors. It vanishes when the momentum transfer approaches zero. 
At leading order, the weak charge is determined by the weak mixing angle, $Q_{\mathrm{W}} \approx 1-4 \sin^2\theta_W$, enabling the inference of the angle itself.

Due to the small weak charge of the proton and the small $Q^2$, the expected asymmetry is very small, $A_{\mathrm{PV}}\approx -27.8\,\mathrm{ppb}$.
A significant challenge is therefore the high statistical demands. 
The goal of P2 is a $2.2\,\%$ measurement of $A_{\mathrm{PV}}$ at the scale $Q^2\approx 0.005\,(\mathrm{GeV}/c)^2$, from which a precision of $0.14\,\%$ can be expected for $\sin^2 \theta_W$. The measurement will be performed using a beam intensity of $150\,\mu\mathrm A$ at an energy of $155\,\mathrm{MeV}$ and a $60\,\mathrm{cm}$ long hydrogen target. $11000\,\mathrm{h}$ data taking time is required to meet the anticipated precision.
%

The precision in extracting $\sin^2 \theta_W$ is presently constrained by our knowledge of the axial form factor of the nucleon. 
To minimize its impact on the systematic error of $\sin^2 \theta_W$, additional measurements will be carried out in elastic electron-proton and electron-deuteron scattering at backward angles. These auxiliary measurements aim to provide further constraints on the axial form factor.

Further studies to determine the weak mixing angle from Parity-Violating Electron Scattering (PVES) off a carbon target are planned as well. In the case of $^{12}$C, the weak mixing angle is sensitive to different combinations of effective four-fermion couplings. The measurement of weak mixing angles of both proton and $^{12}$C will allow to increase the sensitivity to SM extensions. 

\subsection*{Determination of the neutron-skin thickness of $^{\mathbf{208}}$Pb}\label{sec:P2:nskin}
%
PVES also offers a means to determine the location of neutrons within nuclei with minimal model dependence. 
In Born approximation, $A_{\mathrm{PV}}$ for a spin-zero nucleus is proportional to the ratio of weak $F_{W}$ to charge $F_{\mathrm{ch}}$ form factors, 
$A_{PV}\approx\frac{G_{F}Q^{2}}{4\sqrt 2 \pi \alpha_{\mathrm{em}}}\frac{|Q_{W}|F_{W}(Q^{2})}{ZF_{\mathrm{ch}}(Q^{2})}$,
with the weak and the electric charge $Q_{W}$ and $Z$, respectively \cite{CREX}. 
$F_{W}(Q^{2})$,  which is dominated by the neutron distribution, can be deduced from the measured asymmetry $A_{PV}$ and $F_{\mathrm{ch}}(Q^{2})$, which is well known.
Subsequently, from the weak form factor, one can extract the weak radius, followed by the neutron radius inside the nucleus, and ultimately the neutron-skin thickness which is related to the symmetry energy of the nuclear Equation of State (EoS), which is the very same theoretical framework that is used, for instance, in the modeling of neutron stars.

As neutrons are pushed outwards against surface tension, heavy nuclei are expected to form a neutron-rich skin with many neutrons collecting near the surface.
The thickness of this skin is strongly sensitive to the poorly-known density dependence of the symmetry energy near saturation density.
The determination of the neutron-skin thickness of heavy nuclei has far reaching consequences in many areas of physics including heavy-ion collisions, polarized electron and proton scattering off nuclei, precision tests of the SM using atomic parity violation, and nuclear astrophysics \cite{Thiel_nskins}.

Using the P2 apparatus, the Mainz Radius EXperiment (MREX) aims at measuring the neutron skin of $\rm ^{208}Pb$ with enhanced precision compared to existing PREX \cite{PREX21} und CREX measurements \cite{CREX} to further constrain the symmetry energy of the nuclear Equation of State. 

\section{The beam dump experiment DarkMESA}\label{sec:DarkMESA}
%
The DarkMESA experiment is set to conduct a search for sub-GeV Light Dark Matter, complementing the dark sector investigations at MAGIX, 
which specifically target dark photon searches. Dark photons can potentially be produced in the P2 beam dump by a process analogous to photon Bremsstrahlung and can then decay in DM particle pairs $\bar\chi\chi$ if kinematically allowed, as depicted in Fig.~\ref{fig:VariousDPMeasurements}. A fraction of them scatter off electrons or nuclei in the DarkMESA calorimeter \cite{DM_proc}. Unlike direct search experiments, beam dump experiments offer an advantage as the produced DM particles can carry a substantial amount of kinetic energy. This can lead to a notable recoil within the detector, potentially generating a significant and detectable signal.

DarkMESA can operate simultaneously with P2, capitalizing on the high beam intensity and extended measurement campaigns. 
Positioned $23\,\mathrm{m}$ behind the beam dump, the DarkMESA detector will benefit from shielding provided by the walls of the accelerator hall, protecting it from escaping neutrons or other SM particles. The detector consists of two main components: an electromagnetic calorimeter array for detecting signals produced by LDM, and a veto system designed to reject background \cite{DM_Instrumentation}. The experiment will be performed in different stages. 
Initially, Phase A will utilize a relatively small prototype detector, featuring 5$\times$5 PbF$_2$ Cherenkov detector crystals \cite{DM_CherryResponse} with an active volume of $0.004\,\mathrm{m^3}$. These crystals will be surrounded by a hermetic veto system consisting of two layers of plastic scintillators and $1\,\mathrm{cm}$ of lead shielding. Moving to Phase B, the calorimeter will be extended to 30$\times$30 PbF$_2$ and 32$\times$32 SF5 lead-glass blocks, totaling an active volume of $0.7\,\mathrm{m^3}$. In the subsequent Phase C, the active volume will be expanded either through additional lead-glass blocks or by adopting new technologies such as a time-projection chamber or a liquid scintillator detector.

\begin{figure}[h]
    \centering
    \sidecaption
    \includegraphics[width=0.5\columnwidth]{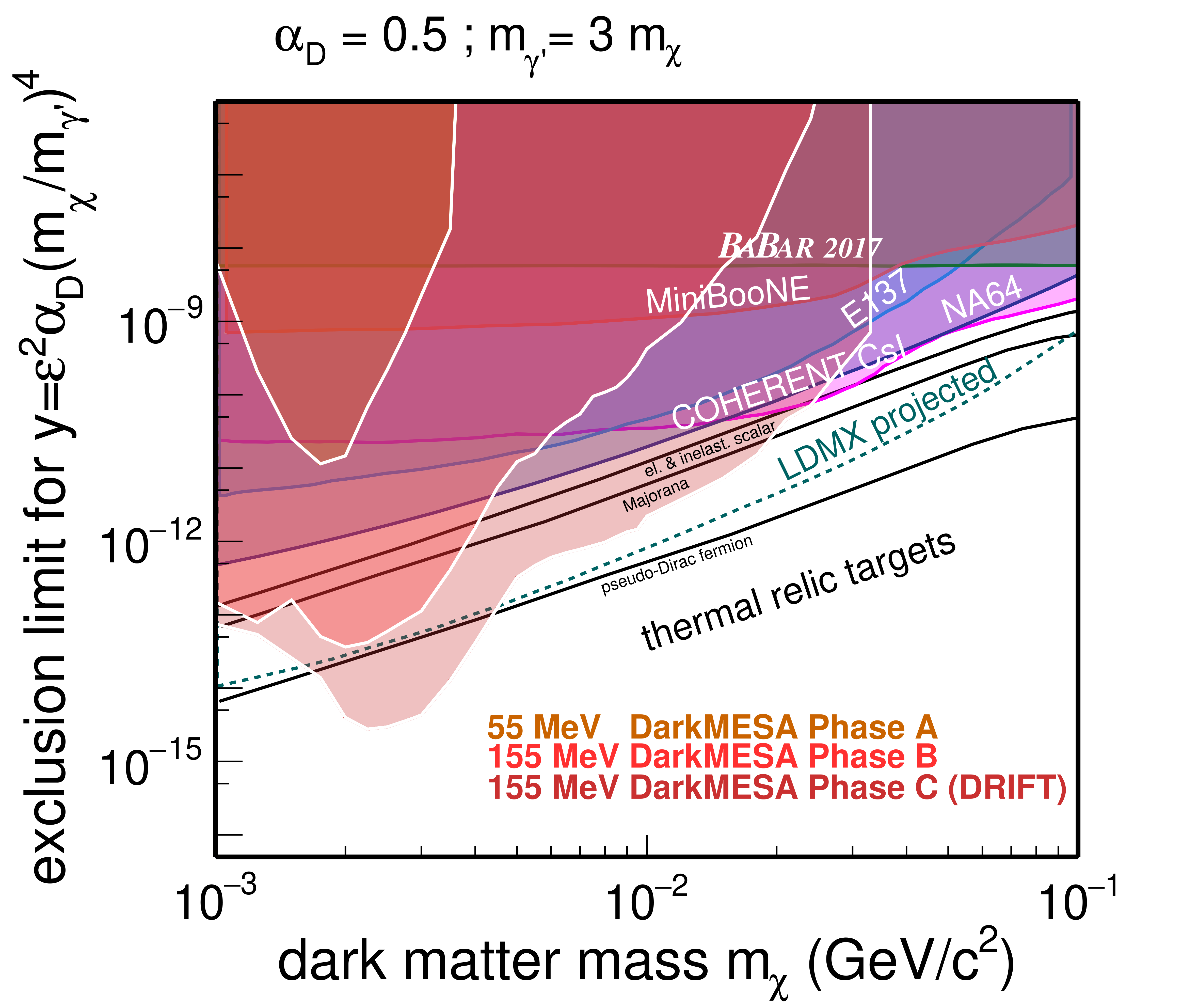} 
    \caption{
        Projected 90\% C.L. exclusion limits for the three proposed phases of the DarkMESA experiment in comparison with current limits. For the Phase C of DarkMESA, a detector based on a time-projection chamber technology (DRIFT) has been assumed. 
        The limits are represented in terms of the yield variable $y$ (natural dimensionless interaction strength) and the DM mass $m_{\chi}$.
        Indicated are also relic density limits, which correspond to predictions assuming a thermal origin of the DM.
        }
        \label{fig:DM_Exclusion}
    \end{figure}

For the precision envisioned by P2, as many as $5 \cdot  10^{22}$ Electrons on Target (EoT) will be needed. The majority of these electrons will be stopped in the beam dump of the experiment. Such a high number of electrons allows for a competitive search for LDM particles. Figure~\ref{fig:DM_Exclusion} depicts the simulation result for the sensitivity of DarkMESA.
%

\section*{Acknowledgments}

This work was supported in part by 
the PRISMA$^+$ (Precision Physics, Fundamental Interactions and Structure of Matter) Cluster of Excellence,
the Deutsche Forschungsgemeinschaft (DFG, German Research Foundation) through the Collaborative Research Center 1044,
the Federal State of Rhineland-Palatinate,
and the European Union's Horizon 2020 research and innovation programme, project STRONG2020, under grant agreement No 824093.
\newpage

\end{document}